\documentclass{article}
\usepackage[a4paper, margin=2.5cm]{geometry}
\usepackage{graphicx}
\usepackage{multirow}
\usepackage{amsmath}
\usepackage{amssymb}
\usepackage{xcolor}
\usepackage{authblk}
\usepackage{cite}
\usepackage[colorlinks,citecolor=blue,linkcolor=blue]{hyperref}                
\newcommand{\rmd}{{\mathrm d}}
\newcommand{\rmi}{{\mathrm i}}
\title{The Jost function and Siegert pseudostates from $R$-matrix calculations at complex wavenumbers} 
\author{
P. Vaandrager, J. Dohet-Eraly and J-M. Sparenberg 
}
\date{%
    \textit{\small Nuclear Physics and Quantum Physics, 
C.P. 229, \\
 Université libre de Bruxelles (ULB), B 1050 Brussels, Belgium} 
  \\[2ex]%
    \today
}

\begin{document}
\maketitle

\begin{abstract}
The single-channel Jost function is calculated with the computational $R$-matrix on a Lagrange-Jacobi mesh, in order to study its behaviour at complex wavenumbers. Three potentials derived from supersymmetric transformations are used to test the accuracy of the method. Each of these potentials, with $s$-wave or $p$-wave bound, resonance or virtual states, has a simple analytical expression for the Jost function, which is compared with the calculated Jost function. Siegert states and Siegert pseudostates are determined by finding the zeros of the calculated Jost function. Poles of the exact Jost function are not present in the calculated Jost function due to the truncation of the potential in the $R$-matrix method. Instead, Siegert pseudostates arise in the vicinity of the missing poles.   
\end{abstract}

\section{Introduction}
The Jost function is a fundamental concept in non-relativistic quantum scattering theory \cite{Taylor,Rakityansky2022}. It contains all the information required for studying a scattering system. In particular, the asymptotic behaviour of a wave function can be given in terms of the Jost function and the scattering matrix ($S$-matrix) can then be defined as a ratio of Jost functions in a simple way. The Jost function thus has simpler analytic properties compared to those of the $S$-matrix. 

Scattering observables such as the phase shift or scattering cross-section can be calculated from the Jost function, and its zeros at specific complex wavenumbers, $k$, correspond exactly with simple poles of the $S$-matrix. Quantum states with such wavenumbers are bound, virtual or resonance states, known as Siegert states collectively \cite{Tolstikhin1997,Tolstikhin1998}. 

In most numerical calculations, zeros of the Jost function appear which do not correspond with Siegert states only. In particular, zeros of the Jost function occur for calculations where the interaction potential of a quantum system is truncated at some radius sufficiently far from the interaction region, where the interaction potential is approximately zero. 
Such zeros of the Jost function correspond with the so-called Siegert pseudostates, which satisfy specific boundary conditions at the radius where the potential is truncated \cite{Tolstikhin1997,Tolstikhin1998}.

Apart from zeros, it is known that the Jost function may have poles for certain $k$ with $\mathrm{Re}(k)=0$ and $\mathrm{Im}(k)<0$. This corresponds with zeros of the $S$-matrix \cite{Taylor}. Interaction potentials with explicit, analytical expressions for the Jost functions can be derived, where such poles are present. However, these poles generally do not appear for numerical calculations of the Jost function, since the interaction potential is truncated in most numerical methods of calculating the Jost function or $S$-matrix. See Refs.~\cite{Rakityansky2022,Rakityansky2013}, for example, where no poles are present in the derived analytic structure of the Jost function, on the assumption that the potential is zero at a large radius. Zeros of the calculated Jost function with wavenumbers in the vicinity of an exact Jost-function pole do arise, instead of a pole \cite{Baye2002}. These are specific Siegert pseudostates which effectively replace the Jost-function pole. 

In this study, numerical calculations are performed using the $R$-matrix formalism, where the configuration space is divided into two regions separated by the channel radius, $a$. This serves as the cut-off radius where the potential is truncated. In the internal region, where $r\le a$, the wave function is expanded on a square-integrable basis. In the external
region where $r \ge a$, it is approximated by its asymptotic behaviour.  

It has been shown that the computational $R$-matrix on a Lagrange mesh can be used to determine phase shifts and cross-sections accurately, as well as bound state energies and resonance parameters, for various potentials \cite{Baye2002,Baye1998,Baye2015,Descouvemont2010}. The method is extended here to calculate the single-channel Jost function at complex $k$.

Short-ranged model potentials with known resonances and bound or virtual states, and with simple analytical expressions for the corresponding Jost functions for $\ell=0$ and $\ell=1$, can be constructed from supersymmetric transformations \cite{Baye2014}. These potentials are then used to test the accuracy of the calculated Jost functions determined from the computational $R$-matrix. The behaviour of the calculated Jost functions on the entire complex $k$-plane is explored. In particular, it is used to calculate the Siegert states and Siegert pseudostates for each of the model potentials.

Various procedures for finding Siegert states and Siegert pseudostates with the computational $R$-matrix exist, but there is little comparison in the accuracy of the methods. The results obtained by finding the zeros of the Jost function are compared with the method from Refs.~\cite{Tolstikhin1997,Tolstikhin1998,Baye2002}, which can only be applied to scattering problems where $\ell =0$.          

In the next section, the Jost function, Siegert states and Siegert pseudostates are discussed. Section~\ref{section_R_matrix} includes an overview of the $R$-matrix on a Lagrange mesh and gives the method of calculating the Jost function from the $R$-matrix. The test potentials are discussed in Section~\ref{section_potentials}. In the results section, the calculated Jost functions are compared with the exact Jost functions. Calculated wavenumbers of the Siegert states and Siegert pseudostates are given and compared with ones obtained by other methods, where possible. The conclusion follows.  
   
\section{The Jost function, Siegert states and Siegert pseudostates}
\subsection{The Jost function and Siegert states}
The radial Schr\"{o}dinger equation is given by 
\begin{equation} \label{SE_simple}
 H_{\ell}  u_{\ell}  = E  u_{\ell} 
\end{equation}
with the two-body Hamiltonian given by:
\begin{equation} \label{Hamiltonian}
H_{\ell} = T_0 + T_{\ell} + V_S(r) + V_{C}(r) ,
\end{equation} 
where $V_S(r)$ is the short-ranged radial potential, which is finite at the origin and which goes to zero faster than $r^{-2}$ as $r\rightarrow \infty$. The Coulomb potential is given by
\begin{equation}
V_{C}(r) = \frac{e^2 Z_1 Z_2}{r}
\end{equation}
and the kinetic operators by
\begin{equation}
T_0 = -\frac{\hbar^2}{2\mu} \frac{d^2}{dr^2} \mbox{ and } T_{\ell} = \frac{\hbar^2}{2\mu}  \frac{\ell(\ell+1)}{r^2} ,
\end{equation}
where $\mu$ is the reduced mass of the two-body system. The square of the channel momentum and Sommerfeld parameter are given as follows, respectively:
\begin{equation}
k^2 = \frac{2\mu E}{\hbar^2} \mbox{ and } \eta = \frac{\mu e^2 Z_1 Z_2}{k\hbar^2}.
\end{equation}
 
The \textit{regular} solution, $\phi_{\ell}(k,r)$, of eq.~(\ref{SE_simple}) is defined by its behaviour at the origin, but the convention for its normalisation differs in some texts. Here it is chosen to behave exactly like the regular Coulomb function $F_{\ell}(kr,\eta)$ \cite{NIST}:
\begin{equation} \label{phi-r_0}
\phi_\ell(k,r)  \xrightarrow[r \to 0]{} F_{\ell}(kr,\eta)  \xrightarrow[r \to 0]{}
C_{\ell}(\eta) (kr)^{\ell+1},
\end{equation}   
where the Coulomb barrier factor is given by
\begin{equation}
C_{\ell}(\eta) = \frac{2^{\ell} e^{-\pi \eta/2}}{\Gamma(2\ell + 2)} 
\vert {\Gamma(\ell+1 \pm i\eta)} \vert.
\end{equation}
When there are no Coulomb interactions, $\eta = 0$ and $C_{\ell}(0) = 1/(2\ell+1)!!$. The normalisation of the regular solution used here then becomes identical to that used in Ref.~\cite{Rakityansky2022}, which is chosen to be the same for Coulomb- and non-Coulomb interactions.
 
The single-channel Jost function is defined as the energy-dependent amplitudes of the incoming and outgoing spherical waves of the regular solution to the
radial wave-equation, at the limit far from the interaction region \cite{Taylor,Rakityansky2022}. The asymptotic behaviour of the regular solution is given as follows \cite{Rakityansky1996}:
\begin{equation} \label{phi_jost_r-inf}
\phi_\ell(k,r) 
\xrightarrow[r \to \infty]{} 
 \frac{i}{2} 
 \left[
f_{\ell}(k)  I_{\ell}(kr,\eta) - f_{\ell}(-k)  O_{\ell}(kr,\eta) \right],
\end{equation}
where $f_{\ell}(k)$ is the Jost function. The functions $I_{\ell}(kr,\eta)$ and $O_{\ell}(kr,\eta)$ correspond with the incoming and outgoing spherical waves, respectively, and  are given in terms of the regular and irregular Coulomb functions as follows:
\begin{equation}
I_{\ell}(kr,\eta) = G_{\ell}(kr,\eta) - iF_{\ell}(kr,\eta) \mbox{ and }
O_{\ell}(kr,\eta) = G_{\ell}(kr,\eta) + iF_{\ell}(kr,\eta). 
\end{equation}
They have the following useful symmetry relation, which can be proven from the properties of the Coulomb functions given in Ref.~\cite{NIST} or, in greater detail, in Ref.~\cite{Gaspard2018}:
\begin{equation} \label{I_O_k_relation}
I^*_{\ell}(-kr,-\eta) = (-1)^{\ell} e^{-\pi\eta} O_{\ell} (kr,\eta).
\end{equation}

The scattering matrix, $S_{\ell}(k)$, is defined in terms of the Jost function by the following \cite{Rakityansky2022}:
\begin{equation} \label{S_k}
S_{\ell}(k) =  \frac{f_{\ell}(-k)}{f_{\ell}(k)}.
\end{equation}
The Jost function has the symmetry property \cite{Rakityansky2022}:
\begin{equation}
 f_{\ell}(-k) = f_{\ell}^*(k^*).
\end{equation}
Siegert states are defined as complex energy eigenstates of eq.~(\ref{SE_simple}). The boundary conditions of the regular solution for the Siegert states are given by eq.~(\ref{phi-r_0}) at the origin, and by the following relation for large $r$ \cite{Tolstikhin1997,Tolstikhin1998}: 
\begin{equation} \label{siegert_cond_inf}
\frac{d}{dr} \phi(k,r \rightarrow \infty) = ik \phi(k,r \rightarrow \infty).
\end{equation}
Using eq.~(\ref{phi_jost_r-inf}) for the regular solution where~$r \rightarrow \infty$, it can be deduced that the condition only holds for states with wavenumber, $k$, where the Jost function, $f_{\ell}(k)$, is zero, which implies a purely outgoing wave function.  This is the case for bound, resonance and virtual states \cite{Taylor}, and of course corresponds with poles of the $S$-matrix by eq.~(\ref{S_k}). 
Bound states occur for~$\mathrm{Re}(k) = 0$,~$\mathrm{Im}(k)>0 $; virtual states for~$\mathrm{Re}(k) = 0$, $\mathrm{Im}(k) < 0 $; resonance states for $\mathrm{Re}(k) > 0, \mathrm{Im}(k) < 0 $ and mirror-resonances symmetrical to physical resonances for~$\mathrm{Re}(k)<0$, $\mathrm{Im}(k) < 0 $ \cite{Taylor}. 

The poles of the $S$-matrix that occur for zeros of the Jost function, $f_{\ell}(k)$, are the ``true’’ poles of the $S$-matrix and are the Siegert states. However, poles of the $S$-matrix also correspond with \textit{poles} of $f_{\ell}(-k)$, the numerator in eq.~(\ref{S_k}). These are known as the false poles of the $S$-matrix \cite{Rakityansky2022}, and do not have clear physical meaning like true $S$-matrix poles. See Ref. \cite{Rakityansky2022} for further details. 

It is important to note that the Jost function as defined in Ref.~\cite{Taylor} and most other texts, is normalised by the behaviour of eq.~(\ref{phi-r_0}) at the origin. Different normalisations result in a Jost function differing by an energy-dependent factor. This factor has no impact on the $S$-matrix, since it cancels out in eq.~(\ref{S_k}) for real scattering energies, where phase shifts or cross-sections are calculated. Or, when the zeros of the Jost function are determined to locate Siegert states, the energy-dependent factor (which is finite) does not affect the search for zeros. See Ref.~\cite{Rakityansky2022} for further details.

\subsection{Truncation of the potential and Siegert pseudostates}
In most numerical calculations, at some $r=a$ the interaction potential is truncated: $V( r\ge a) =0$. Consequently, the Siegert boundary condition for the regular solution is used instead of eq.~(\ref{siegert_cond_inf}):
\begin{equation} \label{siegert_cond}
\frac{d\phi(k,a)}{dr}  = ik \phi(k,a).
\end{equation} 
This approximation still gives accurate results when calculating physical quantities of interest, in particular the Siegert states and phase shifts, provided $a$ is chosen large enough. A consequence of its implementation is the occurrence of the so-called Siegert pseudostates, which are also complex energy eigenvalues of (\ref{SE_simple}), but do not have the same strong physical meaning as the Siegert states. The Siegert pseudostates can occur for positive or negative $\mathrm{Re}(k)$ and $\mathrm{Im}(k) < 0 $, and thus resemble wide resonances.  See Refs.~\cite{Tolstikhin1997,Tolstikhin1998} for a detailed discussion. The Siegert pseudostates are, furthermore, related to exact Jost-function poles.  

The Jost function is mostly entire in $k$, but it may have poles in a certain region of the complex $k$-plane, specifically along the negative imaginary $k$-axis \cite{Taylor}. However, the analytic structure of the Jost function for scattering involving short-ranged and Coulomb interactions is explicitly given in Refs.~\cite{Rakityansky2013,Rakityansky2022}, and possible poles on the imaginary axis of $k$ between $-i\infty$ and $0$ are not accounted for. This is because the Jost function is only analytic on the whole $k$-plane for truncated potentials, apart from further possible poles at $k=0$. 

With the computational $R$-matrix method, the potential is also truncated at some $r=a$, which should result in a Jost function which is analytic in $k$. All the potentials used in this study have exact Jost-function poles at certain negative imaginary values of $k$. However, the calculated Jost function has a finite value dependent on the choice of $a$ at the poles of the exact Jost function. The poles of the exact Jost function manifest in the calculated Jost function by the behaviour of the Siegert pseudostates, which are distributed around the actual pole, as will be seen in Section~\ref{results}.

Since poles of the calculated Jost function (and so $f(-k)$) do not occur and are replaced by Siegert pseudostates, false poles of the calculated $S$-matrix do not occur either, but the corresponding Siegert pseudostates must appear as zeros of the $S$-matrix.

From eq.~(\ref{S_k}), it is clear that each Jost-function zero is an $S$-matrix pole and each Jost-function pole is an $S$-matrix zero. Furthermore, for each $S$-matrix pole at a specific $k$, an $S$-matrix zero will occur at $-k$. When considering Siegert states, Siegert pseudostates and false poles, this makes the analytic structure of the $S$-matrix much more complicated than the analytic structure of Jost function, which is why it is preferred to study the Jost function.  

\section{The $R$-matrix method} \label{section_R_matrix}
\subsection{$R$-matrix formalism}
A comprehensive derivation and explanation of the $R$-matrix method is given in Refs. \cite{Descouvemont2010,Burke}, for example. It must firstly be noted that there are, in fact, two approaches in determining the $R$-matrix for a scattering problem: by fitting the $R$-matrix to experimental data, known as the \textit{phenomenological} $R$-matrix method, or by calculating the $R$-matrix from a model potential, known as the \textit{computational} $R$-matrix method. We restrict ourselves to determine the single-channel Jost function from the computational $R$-matrix. 

In the $R$-matrix method, the configuration space is divided into an internal region and an external region, with a boundary at the channel radius, $r=a$, beyond which the potential is approximately zero: in other words, the potential is truncated. In the internal region with $r \le a$, the wave function is expanded on some finite basis of $N$ linearly independent, square-integrable basis functions:
\begin{equation} \label{u_int}
u_{\ell}(k,r) = \sum_{i = 1}^N c_i \varphi_i(r), \qquad 0 \le r \le a,
\end{equation}
which disappear at $r=0$ and satisfy arbitrary boundary conditions at $r=a$. 

In the space of the basis functions, the Hamiltonian operator, $H_{\ell}$, is not Hermitian nor does it admit a complete set of eigenvectors, since the basis functions have arbitrary boundary conditions at $r=a$. To overcome this, the Bloch operator is introduced, defined by: 
\begin{equation} \label{Bloch_def}
\mathcal{L}(B) = \frac{\hbar^2}{2 \mu}
\delta(r-a) \left(
\frac{d}{dr} - \frac{B}{r}
\right),
\end{equation}
where $B$ is known as the boundary parameter, which can be chosen arbitrarily.
It can be shown that the operator, $\left[H_{\ell}+\mathcal{L}(B) \right]$, is Hermitian for real $B$ in the space of the basis functions \cite{Burke}. This implies that it does admit a complete set of eigenvectors, and is thus diagonalisable. However, there are instances where $B$ is chosen to be complex, in which case the operator, $\left[H_{\ell}+\mathcal{L}(B) \right]$, can no longer be Hermitian. It can, however, be shown that it still admits a complete set of $N$ eigenvectors, $v_{n \ell}$, with eigenvalues, $E_{n\ell}$ for any complex $B$, in the space of the basis functions. It is still diagonalisable, which is important later. 

Because of the Bloch operator, and at the limit of infinite basis, the
eigenfunctions of $\left[H_{\ell}+\mathcal{L}(B) \right]$ have a fixed logarithmic derivative at the boundary $r=a$, the value of which is proportional to $B$. This is not the case for a finite basis, for which non-converged eigenfunctions may display different boundary behaviours. It might thus be tempting to enforce the boundary condition by choosing basis functions, $\varphi_i(r)$ that satisfy it. This turns out to be a bad practice leading to numerical inaccuracies \cite{Descouvemont2010}, because physical wave functions at arbitrary energies have varying boundary conditions which are better covered by a basis without fixed logarithmic derivative at the boundary.

Using the Bloch operator, eq.~(\ref{SE_simple}) can be written as follows:
\begin{equation} \label{introd_G}
\left[ H_{\ell}- E + \mathcal{L}(B)  \right] 
u_{\ell}(k,r) = \mathcal{L}(B) u_{\ell}(k,r),
\end{equation}

This is a linear differential equation where the Green's function, $G_{\ell}(r,r')$ of the differential operator, $\left[H_{\ell}-E+\mathcal{L}(B) \right]$, must satisfy the following relation, per definition:
\begin{equation}
\left[H_{\ell}-E+\mathcal{L}(B) \right] G_{\ell}(r,r')
= \delta(r-r').
\end{equation}

The Green's function can be approximated in the internal region by an expansion over a finite set of the basis functions:
\begin{equation} \label{approx_green}
G_{\ell}(r,r') = \sum_{i,j=1}^N \varphi_i(r) 
\left[C(E,B)^{-1} \right]_{ij}
\varphi_j(r') \mbox{ with } 0 \le r,r' \le a 
\end{equation}
and with the matrix elements of $C$ given by:
\begin{equation} \label{C_def}
 C_{ij}(E,B) =
 \int_0^a 
\varphi_i(r) 
\left[ H_{\ell} - E + \mathcal{L}(B)  \right]
\varphi_j(r)  
dr .
\end{equation}
The expression for the Green's function would be exact, if it were expanded over a complete basis. The wave function can be determined with the Green's function \cite{Descouvemont2010}:
\begin{equation} \label{u_G_L}
u_{\ell}(k,r) = \int_0^{\infty} G_{\ell}(r,r') \mathcal{L}(B) u_{\ell}(r') dr'.
\end{equation}
Using eq.~(\ref{Bloch_def}) as an operator of the variable $r'$ in eq.~(\ref{u_G_L}), the following is obtained:
\begin{eqnarray}
u_{\ell}(k,r) 
&=& 
\left[ \frac{d u_{\ell}(k,a)}{dr}
- \frac{B}{a} u_{\ell}(k,a) \right]
\frac{\hbar^2}{2 \mu}
G_{\ell}(r,a) \nonumber \\
&=&
\left[ \frac{d u_{\ell}(k,a)}{dr}
- \frac{B}{a} u_{\ell}(k,a) \right]
\frac{\hbar^2}{2 \mu}
\sum_{i,j=1}^N \varphi_i(r) 
\left[C(E,B)^{-1} \right]_{ij}
\varphi_j(a),\qquad 0 \le r \le a . \label{u_R_init}
\end{eqnarray}
This expression can be simplified by introducing the $R$-matrix, which is
defined at the boundary, $r = a$, as follows:
\begin{equation} \label{u_R_def}
u_{\ell}(k,a) = \left[ a \frac{du_{\ell}(k,a)}{dr} - B u_{\ell}(k,a) \right] 
R_{\ell}(E,B).
\end{equation}
Comparison with eq.~(\ref{u_R_init}) gives the relation between the $R$-matrix and the Green's function:
\begin{equation}
R_{\ell}(E,B) = \frac{\hbar^2}{2 \mu a}
G_{\ell}(a,a).
\end{equation}

By eq.~(\ref{approx_green}), the $R$-matrix is then given by:
\begin{equation} \label{R_inverse_mat}
R_{\ell}(E,B) = \frac{\hbar^2}{2 \mu a} \sum_{i,j=1}^N \varphi_i(a) 
\left[C(E,B)^{-1} \right]_{ij}
\varphi_j(a).
\end{equation}

Since the operator, $\left[H_{\ell}+\mathcal{L}(B) \right]$, is diagonalisable and admits a complete set of eigenvectors for any $B$ in the space of the basis functions, the inverse of matrix $C(E,B)$ can be expanded in the finite spectral decomposition of $N$ eigenvalues:
\begin{equation} \label{spectral_C}
\left[ C(E,B)^{-1} \right]_{ij} =  \sum_{n=1}^N \frac{v_{n\ell}v_{n\ell}^T}{E_{n\ell}-E}.
\end{equation}  
The following familiar expression for the $R$-matrix is then obtained:
\begin{equation} \label{spectral_R}
R_{\ell} (E,B) =  \sum_{n=1}^N \frac{\gamma_{n\ell}\gamma_{n\ell}^T}{E_{n\ell}-E},
\end{equation}  
with
\begin{equation}
\gamma_{n\ell} = \sqrt{\frac{\hbar^2}{2 \mu a}} \sum_{i=1}^N v_{n\ell,i}\varphi_i(a),
\end{equation}
where $v_{n\ell,i}$ is the $i^{\mathrm{th}}$ component of $v_{n\ell}$.

The $R$-matrix can then be calculated in two ways: firstly, by finding the inverse of $C(E,B)$ of eq.~(\ref{C_def}) at a specific $E$ and using the result in eq.~(\ref{R_inverse_mat}). Or, secondly, by finding the eigenvalues, $E_{n\ell}$, and eigenvectors, $v_{n\ell}$, of $C(0,B)$ and using eq.~(\ref{spectral_R}). The second method is preferred in this study, as the $R$-matrix for all complex $E$ (and thus $k$) is obtained by calculating eigenvalues and -vectors only once. Both methods were, however, attempted and give identical numerical accuracy, with insignificant difference in computational time.  

The $R$-matrix is entire in $E$, apart from the $N$ simple poles at $E=E_{n\ell}$. For real $B$, these poles will be real, and for complex $B$, these poles will also be complex.

Using eq.~(\ref{u_R_def}), the wave function in the interior is given in terms of the $R$-matrix and the wave function at the boundary:
\begin{equation} \label{u_R_next}
u_{\ell}(k,r) 
=
\frac{u_{\ell}(k,a)}{R_{\ell}(E,B)}
\frac{\hbar^2}{2 \mu a}
\sum_{i,j=1}^N \varphi_i(r) 
\left[C(E,B)^{-1} \right]_{ij}
\varphi_j(a),\qquad 0 \le r \le a . 
\end{equation}
 
The normalisation of the wave function in the interior will be fixed by the normalisation of the wave function at the boundary, which will be approximated by the asymptotic behaviour of the wave function \cite{Taylor}:
\begin{equation} \label{asymptote_u}
u_{\ell}(k,r)
\xrightarrow[r \to \infty]{} 
  I_{\ell}(kr,\eta) - S_{\ell}(k)  O_{\ell}(kr,\eta).
\end{equation} 

The wave function and its derivative at the boundary are then given by:
\begin{eqnarray}
u_{\ell}(k,a)
&=& 
  I_{\ell}(ka,\eta) - S_{\ell}(k)  O_{\ell}(ka,\eta), \label{u_at_a} \\
\frac{d}{dr} u_{\ell}(k,a)
&=& 
 \frac{d}{dr} I_{\ell}(ka,\eta) - S_{\ell}(k) \frac{d}{dr} O_{\ell}(ka,\eta). \label{du_at_a}
\end{eqnarray}

Using eqs.~(\ref{u_at_a}) and (\ref{du_at_a}) in eq.~(\ref{u_R_def}) gives the following approximate expression for the $S$-matrix in terms of the $R$-matrix \cite{Descouvemont2010,Burke}:
\begin{equation} \label{S_R}
S_{\ell}(k) = \frac{I_{\ell}(ka,\eta)}{O_{\ell}(ka,\eta)} 
\frac{1- \left[ L^*_{\ell}(k)-B \right] R_{\ell}(E,B)  }
{1- \left[ L_{\ell}(k)-B \right] R_{\ell}(E,B)  },
\end{equation}
where $L_{\ell}(k)$ is the logarithmic derivative at the channel radius $a$, defined by:
\begin{equation} \label{L_def}
L_{\ell}(k) =  \frac{a}{O_{\ell}(ka,\eta)} \frac{dO_{\ell}(ka,\eta)}{dr} .
\end{equation}

Having established eq.~(\ref{S_R}) for the $S$-matrix, eq.~(\ref{u_at_a}) can now be used to fix the absolute normalisation of the wave function in the interior region by imposing its continuity at the boundary. Substituting it into eq.~(\ref{u_R_next}) with eq.~(\ref{S_R}) for the $S$-matrix, the following approximate expression for the wave function in the internal region in terms of the $R$-matrix is obtained: 
\begin{equation} \label{u_R_int_f}
u_{\ell}(k,r) 
=
\frac{-ik\hbar^2}{\mu O_{\ell}(ka,\eta)}
\frac{1}{1- \left[ L_{\ell}(k)-B \right] R_{\ell}(E,B)  }
\sum_{i,j=1}^N \varphi_i(r) 
\left[C(E,B)^{-1} \right]_{ij}
\varphi_j(a),\qquad 0 \le r \le a . 
\end{equation}
It should be stressed that this normalization does not guarantee the
continuity of the derivative of the wave function at the boundary, which
is only reached at the infinite basis limit.

As shown in Ref.~\cite{Descouvemont2010}, the wave function and the $S$-matrix are, remarkably, independent of the choice of boundary parameter, $B$. A choice of $B=0$ is usual in the $R$-matrix theory, or another real value which results in a real $R$-matrix in the usual sense. However, a complex choice of $B = L_{\ell}(k)$ is particularly useful. It results in a complex extension of the usual $R$-matrix: see Refs.~\cite{Schneider1981,Descouvemont1990,Ducru2022} for example. The following simple $S$-matrix is obtained if $B = L_{\ell}(k)$: 
\begin{equation} 
S_{\ell}(k) = \frac{I_{\ell}(ka,\eta)}{O_{\ell}(ka,\eta)} 
{
\left\{ 1- \left[ L^*_{\ell}(k)-L_{\ell}(k) \right] R_{\ell}[E,L_{\ell}(k)]
\right\} 
   } .
\end{equation}   
In this case, the complex $R$-matrix poles correspond \textit{exactly} with the $S$-matrix poles, since the factor $I_{\ell}(ka,\eta)/O_{\ell}(ka,\eta)$ is finite for all $k$. It must be stressed that the $S$-matrix poles are the same for any choice of $B$, but the $R$-matrix poles depend on the choice of $B$.  

\subsection{The $R$-matrix and Jost function}
The Jost function calculated with the $R$-matrix method is based on the approximation of the wave function expanded over an appropriate finite basis, $u{_\ell}(k,r)$ given by eq.~(\ref{u_R_int_f}). This wave function and the regular solution, $\phi_\ell(k,r)$, can only differ by an energy-dependent factor proportional to the Jost function. When considering the asymptotic behaviour of $\phi_\ell(k,r)$ and $u{_\ell}(k,r)$,  given in eq.~(\ref{phi_jost_r-inf}) and eq.~(\ref{asymptote_u}) respectively, also using eq.~(\ref{S_k}) for the $S$-matrix, the following relation is obtained, which must hold for all $r$: 
\begin{equation} 
\phi_\ell(k,r) = \frac{i}{2}f_{\ell}(k)    u_{\ell}(k,r).
\end{equation} 

By the behaviour of the regular solution near the origin, eq.~(\ref{phi-r_0}), the Jost function can then be determined with:
\begin{equation}
f_{\ell}(k) \label{f_derive}
 = -2i \lim_{r \rightarrow 0} \frac{F_\ell(kr,\eta)}{u_\ell(k,r)}
 = -2i \lim_{r \rightarrow 0} \frac{C_{\ell}(\eta) (kr)^{\ell+1}}{u_\ell(k,r)}.
\end{equation} 
Using eq.~(\ref{u_R_int_f}), the approximation for $u{_\ell}(k,r)$ in the internal region, results in the following expression for the Jost function:
\begin{equation}\label{f_calc}
f_{\ell}(k) =  C_{\ell}(\eta)
\frac{2 k^{\ell} \mu}{\hbar^2}
O_{\ell}(ka,\eta)
\left[1-(L_{\ell}(k)-B)R_{\ell}(E,B) \right]
\frac{1}{\chi(B)},
\end{equation} 
with
\begin{equation} \label{chi}
\chi(B) =
\sum_{i,j=1}^N 
\left[ \lim_{r\rightarrow 0} \frac{\varphi_i(r)}{r^{\ell+1}} \right] 
\left[ C(E,B)^{-1}\right]_{ij}
 \varphi_j(a).
\end{equation}
To accurately determine the Jost function, calculations need to be performed with an appropriate set of basis functions, which will determine the exact behaviour of the limit in eq.~(\ref{chi}), so that it exists. This choice of basis function is discussed in the next section.

Using eq.~(\ref{I_O_k_relation}), the following expression for $f_{\ell}(-k)$ is obtained: 
\begin{equation}\label{f_out_calc}
f_{\ell}(-k) =  C_{\ell}(\eta)
\frac{2 k^{\ell} \mu}{\hbar^2}
I_{\ell}(ka,\eta)
\left[1-(L^{*}_{\ell}(k)-B)R_{\ell}(E,B) \right]
\frac{1}{\chi(B)}.
\end{equation} 
Substituting eqs.~(\ref{f_calc}) and (\ref{f_out_calc}) into eq.~(\ref{S_k}) recovers eq.~(\ref{S_R}), as expected. 

For the same choice of $B$, the poles of the $R$-matrix and of the function, $\chi(B)$ must be the same, since both are functions of the inverse of $C(E,B)$. By the spectral decomposition given in eq.~(\ref{spectral_C}), the poles of the $R$-matrix and of $\chi(B)$ are simply the eigenvalues, $E_{n\ell}$ of $C(0,B)$.  

Furthermore, the Jost function is completely independent of the choice of $B$, like the $S$-matrix. This is not obvious from eq.~(\ref{f_calc}). It can be proven using eq.~(B3) in Appendix B of Ref.~\cite{Descouvemont2010}, but it also follows from the fact that the wave function used in eq.~(\ref{f_derive}) is independent of $B$ \cite{Descouvemont2010}. For simplicity, all calculations of the Jost function are performed with the simplest choice of $B=0$:
\begin{equation}\label{f_calc_B0}
f_{\ell}(k) =  C_{\ell}(\eta)
\frac{2 k^{\ell} \mu}{\hbar^2}
O_{\ell}(ka,\eta)
\left[1-L_{\ell}(k)R_{\ell}(E,0) \right]
\frac{1}{\chi(0)}.
\end{equation} 

For a choice of $B = L_{\ell}(k)$, the Jost function no longer depends on the $R$-matrix, and it can be calculated with:
\begin{equation}
f_{\ell}(k) =  C_{\ell}(\eta)
\frac{  2 k^\ell  \mu}{\hbar^2}
O_{\ell}(ka,\eta)
\frac{1}{\chi[L_{\ell}(k)]}
. \label{f_calc_Bl}
\end{equation}
In this expression, the zeros of the Jost function are only due to the poles of $\chi[L_{\ell}(k)]$, which are the same as the complex $R$-matrix poles for $B = L_{\ell}(k)$. These are, of course, the $k$ or $E$ that correspond with Siegert states and Siegert pseudostates. Eq.~(\ref{f_calc_Bl}) is less useful for practical calculations, since $\chi[L_{\ell}(k)]$ is numerically more difficult to calculate than $\chi(0)$.  

\subsection{$R$-matrix on a shifted Lagrange-Jacobi mesh}
Performing $R$-matrix calculations on a Lagrange mesh involves calculating the integral in eq.~(\ref{C_def}) using Gauss quadrature. This is an approximation where the definite integral over the function is replaced by a weighted sum of function values at specific points, $r_i$ in the interval $0$ to $a$. Comprehensive details on calculating the $R$-matrix using the Lagrange-Jacobi mesh can be found in Refs. \cite{Baye1998,Baye2015}. Only the main equations are given here. 

The $N$ mesh points, $r_i$, where $i=1,2,...,N$ over the interval $[0,a]$ are obtained by finding the $N$ zeros of the shifted Jacobi polynomial:
\begin{equation}
P^{(0,\beta)}_N \left( \frac{2r_i}{a} - 1 \right) = 0.
\end{equation}             

The corresponding $N$ shifted Lagrange-Jacobi functions regularised by $r/a$ are defined by:
\begin{equation} \label{lagarange-jacobi}
\varphi_i(r) = (-1)^{N-i} \sqrt{\frac{a-r_i}{a r_i}} 
\frac{P^{(0,\beta)}_N \left(\frac{2r}{a}-1 \right) }{r-r_i} 
a^{-\beta/2} r^{\beta/2+1},
\end{equation}  
with
\begin{equation}
\varphi_i(r_j) = \lambda_i \delta_{ij}
\end{equation}    
where $\lambda_i$ are the weights associated with the Gauss quadrature.
For $\beta = 2\ell$, the basis functions have the correct behaviour at the origin. Furthermore, the limit in eq.~(\ref{chi}) exists, and is given as follows:
\begin{equation} \label{lim_f-r}
\lim_{r\rightarrow 0} \frac{\varphi_i(r)}{r^{\ell+1}} = (-1)^{i+1} 
\sqrt{\frac{a-r_i}{ar_i^3}} \frac{(N+2\ell)!}{N!(2\ell)!} \frac{1}{a^{\ell}}.
\end{equation}
Using the Gauss-Jacobi quadrature associated with the mesh, the basis functions are orthonormal:
\begin{equation}
\int_0^a \varphi_i(r) \varphi_j(r) dr   \underset{\mathrm{Gauss}}{=} \delta_{ij}.
\end{equation}
When the Gauss approximation is also used for the potentials terms, the matrix elements of $C(E,B)$ are given explicitly as follows \cite{Baye1998,Baye2015}:
\begin{equation} \label{C_matrix}
C_{ij}(E,B) = T^0_{ij} +\mathcal{L}_{ij}(B)+
\left[
\frac{\hbar^2}{2\mu} \frac{\ell(\ell+1)}{r_i^2}
+V_S(r_i)+V_C(r_i)
-E  
\right]\delta_{ij},
\end{equation}
where the Bloch operator matrix elements, $\mathcal{L}_{ij}(B)$, are given by:
\begin{equation} \label{Bloch_matrix}
\mathcal{L}_{ij}(B) = \mathcal{L}_{ij}(0) - \frac{\hbar^2}{2\mu} \frac{B}{a} \varphi_{i}(a) \varphi_{j}(a).
\end{equation}

The matrix elements $T_{ij}^0 +\mathcal{L}_{ij}(0)$ are given in Ref.~\cite{Baye2015}. They read explicitly, for $i=j$, 
\begin{equation}
T^0_{ij}+\mathcal{L}_{ij}(0)= 
\frac{1}{24r_i(a-r_i)} 
 \left[
4(2N+\beta+1)^2-3\beta^2+8-\frac{\beta^2-4}{r_i}a -\frac{20}{a-r_i}a
\right],
\end{equation}
and, for $i\ne j$,
\begin{eqnarray}
T^0_{ij}+\mathcal{L}_{ij}(0)=
\frac{(-1)^{i-j+1}}{2 \sqrt{r_i r_j(a-r_i)(a-r_j)}}
\qquad \qquad \qquad  \qquad \qquad   \qquad  \qquad \qquad
   \nonumber \\
\times \left[ 
N(N+\beta + 1)+\frac{\beta}{2}+1+ \frac{a r_i+a r_j -2r_i r_j}{(r_i-r_j)^2} 
  -\frac{a}{a-r_i}
  -\frac{a}{a-r_j}  
 \right]. \label{TL_Jacobi_i-not-j}
\end{eqnarray}
For $\ell=0$, the shifted Lagrange-Jacobi mesh becomes exactly the shifted Lagrange-Legendre mesh \cite{Baye1998}. The Lagrange-Legendre mesh was used in Ref.~\cite{Baye2002}, where it was shown that using the Gauss approximation  does not affect the accuracy of the $R$-matrix method.

\subsection{R-matrix methods of calculating Siegert states and pseudostates} \label{section-R-matrix_siegert}
As shown in Ref.~\cite{Baye2002}, the method of Refs.~\cite{Tolstikhin1997,Tolstikhin1998} and the $R$-matrix formalism correspond exactly for $B=L_{\ell}(k)$ defined in eq.~(\ref{L_def}). The Siegert condition, eq.~(\ref{siegert_cond}), is met with such a choice of $B$. As required, a purely outgoing wave function for $r \ge a$ is obtained.  

The Siegert state energies and Siegert pseudostate energies are the eigenvalues, $E_{n\ell}$, of the matrix, $C(0,B)$, of eq.~(\ref{C_matrix}) with $B=L_{\ell}(k_{n\ell})$. Calculating these eigenvalues is not as simple as for a choice of $B=0$, since $E_{n\ell}$ is embedded in $L_{\ell}(k_{n\ell})$ via $k_{n\ell}^2 = 2\mu E_{n\ell}/\hbar^2$. 

For $\ell = 0$ and $\eta = 0$, the external logarithmic derivative is given simply by $L_{0} = ika$. Using eq.~(\ref{C_matrix}) and eq.~(\ref{Bloch_matrix}), the following eigenvalue equation is constructed:
\begin{equation} \label{siegert_method_lis0}
\left[ C_{ij}(0,0) - i \frac{\hbar^2}{2\mu} k_{n0} \varphi_i(a) \varphi_j(a) -\frac{\hbar^2}{2 \mu} k_{n0}^2 \delta_{ij} \right]v_{n0}  
= 0.
\end{equation}  
Solving this for all $k_{n0}$ will give the Siegert states and Siegert pseudostates. This quadratic matrix eigenvalue equation can be written as an easily-solvable generalized eigenvalue problem of double the size \cite{Tolstikhin1998}. This method, that we refer to as the Siegert method, is used in Ref.~\cite{Baye2002} for the Bargmann potential. For $\ell>0$ and $\eta \neq 0$, however, such an algebraic method does not exist. 

Ref.~\cite{Baye2002} suggests using an iterative method, where initially $B=0$. Using eq.~(\ref{C_matrix}), one of the resulting eigenvalues of $C(0,0)$ for the first iteration, $E^{(1)}_{n\ell}$, is then used in $B=L_{\ell}(k_{n\ell}^{(1)})$ with $k_{n\ell}^{(j)}=\pm \sqrt{2\mu E_{n\ell}^j/\hbar^2}$ for $j=1,2,...$. The relevant eigenvalue, $E^{(2)}_{n\ell}$ of $C(0,L_{\ell}(k_{n\ell}^1))$ is then used in the next iteration. The calculation is repeated until the eigenvalue converges for some $j$. This simple scheme is in fact a standard method for calculating bound states with the $R$-matrix formalism and can be used in calculating resonance parameter as well \cite{Descouvemont2010,Schneider1981,Descouvemont1990}. It will be referred to as the iterative $R$-matrix method.   

Since the Siegert condition, eq.~(\ref{siegert_cond}), holds for states where the Jost function is zero, an alternative method for finding the Siegert states and Sigert pseudostates, which we will call the Jost method, simply involves finding the zeros of the calculated Jost function for $B=0$. The Siegert state and Siegert pseudostate energies for the other two methods are eigenvalues of $C(0,L_{\ell}(k))$, with $B=L_{\ell}(k)$. By the spectral decomposition, eq.~(\ref{spectral_C}), these energies correspond with poles of $C(E,L_{\ell}(k))^{-1}$. By eq.~(\ref{f_calc_Bl}), these energies or corresponding $k$ result in zeros of the Jost function for $B=L_{\ell}(k)$, since the poles of $C (E,L_{\ell}(k))^{-1}$ lead to poles of $\chi(L_{\ell}(k))$. However, the Jost function is independent of $B$ and it will have the same zeros for the choice of $B=0$. The Jost method is thus equivalent to the other two methods, and it can be used for any $\ell$ and $\eta$, unlike the Siegert method. Finding poles of the $S$-matrix is, of course, equivalent to finding zeros of the Jost function. Finding poles of the $S$-matrix is done in Ref.~\cite{Schneider1981}, for example, to locate Siegert states.      
       
\section{Test potentials} \label{section_potentials}
Three single-channel potentials with known, exact expressions for the Jost functions will be considered, to test the accuracy of eq.~(\ref{f_calc_B0}). For simplicity, all three potentials will not include Coulomb interactions ($\eta=0$), and $\hbar = \mu = 1$. 

\subsection{Potential 1 where $\ell=0$ with bound or virtual state}
The first potential that will be used is the Eckart potential \cite{Eckart1930}, which falls under a class of potentials derived by Bargmann \cite{Bargmann1949} and is thus often referred to as the Bargmann potential instead. It is derived from supersymmetric transformations in Ref.~\cite{Baye2014}. In its simplified form, it is given by:
\begin{equation}
V(r) =  -4 \kappa_0^2 \frac{\beta_V e^{-2 \kappa_0 r}}
{\left[1+\beta_Ve^{-2 \kappa_0 r} \right]^2},\qquad
\beta_V=\frac{\kappa_0+\kappa_1}{\kappa_0-\kappa_1}>1,  \label{V1}
\end{equation}
where $\kappa_0$ and $\kappa_1$ are real quantities, with $\kappa_0>0$. In order for the potential not to have a singularity, it is required that $\kappa_0 > \kappa_1$. The value of the potential at the origin is: $ V(0) = 2 (\kappa_1^2-\kappa_0^2)<0 $. The corresponding Jost function is given by:
\begin{equation} \label{exact_Jost_V1}
f_{0}(k) = \frac{k-i\kappa_1}{k+i\kappa_0}.
\end{equation}
A bound or virtual state occurs for $k=i\kappa_1$ (a zero of the Jost function), with $\kappa_1>0$ for a bound state and $\kappa_1<0$ for a virtual state. A Jost-function pole occurs at $k=-i\kappa_0$, where it is required that $\kappa_0 > 0$. 

This potential is identical to the one used in Ref.~\cite{Baye2002}, with $\kappa_0 = b $ and $\kappa_1 = -c $. The exact expression for the phase shift for this potential is given by eq.~(4.16) of Ref.~\cite{Baye2014}.

\subsection{Potential 2 where $\ell=0$ with resonance}
For the second potential, we will use eqs.(4.47)-(4.51) of Ref.~\cite{Baye2014}. There was an error in the signs of $\zeta_i$, which has been corrected here:
\begin{equation}
V(r)  =  \frac{\left(\kappa_0^2-\kappa_1^2\right)
\left[\kappa_1^2\sinh^2\left(\kappa_0 r+\zeta_0\right)
	-\kappa_0^2\sinh^2\left(\kappa_1 r+\zeta_1 \right)\right]}
{\left[\kappa_1\sinh\left(\kappa_0 r+\zeta_0\right)\cosh\left(\kappa_1 r+
\zeta_1 \right)-
	\kappa_0\sinh\left(\kappa_1 r+\zeta_1\right)\cosh\left(\kappa_0 r+
\zeta_0\right)\right]^2},  \label{V2}
\end{equation}
where the real positive constants, $\zeta_i$ are defined by eq.~(4.46) of Ref.~\cite{Baye2014}, where a further error in the first expression has been corrected:
\begin{equation} \label{sigma_sign}
\zeta_i = \mathrm{arctanh}\dfrac{\kappa_i}{\alpha} + \mathrm{arctanh}\dfrac{\kappa_i}{\alpha^*}
= \mathrm{arctanh} \dfrac{ 2 \alpha_R \kappa_i}{\kappa_i^2 + |\alpha|^2}.
\end{equation}
The parameters $\kappa_0$ and $\kappa_1$ are real and positive while $\alpha = \alpha_R + i \alpha_I$ is complex and $\alpha_R$ and $\alpha_I$ are real.    

Assuming $\kappa_1 > \kappa_0$, the potential is not singular 
if $\kappa_0 \tanh \zeta_1 > \kappa_1 \tanh \zeta_0$. The value of the potential at the origin cannot be given in terms of the parameters with such a simple expression as for potential 1. 

The corresponding exact Jost function reads:
\begin{equation} \label{exact_Jost_V2}
f_0(k) = \frac{(k+\rmi\alpha)(k+\rmi\alpha^*)}{(k+\rmi\kappa_0)(k+\rmi\kappa_1)}\,.
\end{equation}
There is thus a resonance at $k=-i\alpha$ (with a mirror-resonance at $k=-i\alpha^*$) and two Jost-function poles at $k=-i\kappa_0$ and $k=-i\kappa_1$. The exact expression for the phase shift for this potential is given by eq.~(4.51) of Ref.~\cite{Baye2014}.

\subsection{Potential 3 where $\ell=1$ with resonance}
This potential is not given in Ref.~\cite{Baye2014} and will therefore be discussed in more detail. It is an extension of the second potential, but with $\ell=1$. It is also derived from supersymmetric transformations and has the same expression for the exact Jost function as potential 2. The phase shifts for both potentials are consequently also the same. The potential is given by:
\begin{eqnarray}
V(r) 
& = &  -2 \frac{\rmd^2}{\rmd r^2}
\ln \mathrm{W}\left[\left(1+\frac{1}{\alpha r}\right) e^{-\alpha r},
\left(1+\frac{1}{\alpha^* r}\right) e^{-\alpha^* r}, \right. \nonumber \\
 & & \left.  \cosh(\kappa_1 r)-\frac{\sinh(\kappa_1 r)}{\kappa_1 r},
\cosh(\kappa_2 r)-\frac{\sinh(\kappa_2 r)}{\kappa_2 r}\right]
 \label{V3i}
 \\
  & = &  -  \frac{\rmd^2}{\rmd r^2} \ln
\mathrm{W}[r, e^{-\alpha r}, e^{-\alpha^* r},
\sinh(\kappa_0 r), \sinh(\kappa_1 r)]  -\frac{1}{r^2}  \label{V3ii} \\
& = & -  \frac{\rmd^2}{\rmd r^2} \ln
\mathrm{W}\left[r + \frac{2\alpha_R}{|\alpha| ^2},
\sinh\left(\kappa_0 r -\zeta_0\right),
\sinh\left(\kappa_1 r -\zeta_1\right)\right] -\frac{1}{r^2}.  \label{V3}
\end{eqnarray}
Note that the centrifugal term, $T_{1}=1/r^2$, is incorporated in the Wronskian, $W(r)$, in eqs.~(\ref{V3ii}) and (\ref{V3}) and is thus subtracted in both these expressions. Furthermore, the exponentials in the Wronskian allows one to reduce the order between eq.~(\ref{V3ii}) and eq.~(\ref{V3}).
For this potential to be short-ranged (except for the centrifugal term) and for the corresponding phase shift to satisfy the corresponding effective-range expansion, the parameters have to satisfy the additional condition:
\begin{equation} \label{V3_cond}
0  =  \frac{1}{\kappa_0}+\frac{1}{\kappa_1}-\frac{1}{\alpha}-\frac{1}{\alpha^*}
= \frac{1}{\kappa_0}+\frac{1}{\kappa_1}-\frac{2\alpha_R}{|\alpha| ^2}.
\end{equation}
At small $r$, numerical instabilities leading to quasi singularities occur, when using any of the three equivalent expressions, eqs.~(\ref{V3i})-(\ref{V3}). In eq.~(\ref{V3i}), this is due to denominators in various terms containing a factor $r$. In eqs.~(\ref{V3ii}) and (\ref{V3}), it is due to the centrifugal term being embedded in the Wronskian. However, it is known that the potential is finite at $r=0$ if the condition, eq.~(\ref{V3_cond}) is met.

A possible way of dealing with this numerical instability, which is applied in this study, is to use eq.~(\ref{V3}) and first to apply the sum-of-angles relation for $\sinh$:
\begin{equation}
\sinh\left(\kappa_i r -\zeta_i\right) = \sinh\left(\kappa_i r\right) \cosh\left(\zeta_i\right)
- \cosh\left(\kappa_i r\right) \sinh\left(\zeta_i\right),
\end{equation}
and then to approximate the hyperbolic trigonometric functions with the first $m$ terms of their Taylor expansions:
\begin{equation}
\sinh(\kappa_i r) \approx  \sum_{n=0}^{m-1} \frac{(\kappa_i r)^{2n+1}}{(2n+1)!} \mbox{ and }
\cosh(\kappa_i r) \approx  \sum_{n=0}^{m-1} \frac{(\kappa_i r)^{2n}}{(2n)!}.
\end{equation}
This approximation for the potential was used for $r<0.05$, with $m=6$. In particular, the value of the potential at the origin can be determined, but there is no simple expression for $V(0)$.  
     
\section{Results} \label{results}
\subsection{Potential 1} 
The $\ell=0$ potential with bound state, eq.~(\ref{V1}), is used with parameters $\kappa_0=2$ and $\kappa_1=1$, which corresponds exactly with the Bargmann potential used in Ref.~\cite{Baye2002}. Since the Lagrange-Jacobi mesh with $\ell=0$ reduces to the Lagrange-Legendre mesh, the phase shifts reported in Ref.~\cite{Baye2002} could be reproduced for $a=5$ and $N=25$.   

In our Jost-function calculations, the same channel radius of $a=5$ is chosen, since the potential is approximately zero for $r \ge 5$. For a reference value of $k=0.5$, the number of mesh points, $N$ is steadily increased until values for the calculated Jost function converge. This occurs at $N=40$, with an accuracy of $8$ digits. For different $k$, convergence of the calculated Jost function occurs for different $N$, but for simplicity all calculated values are given for the same choice of $a=5$ and $N=40$, unless otherwise stated.

\begin{figure}[htbp]
\centerline{\includegraphics[width=\textwidth]{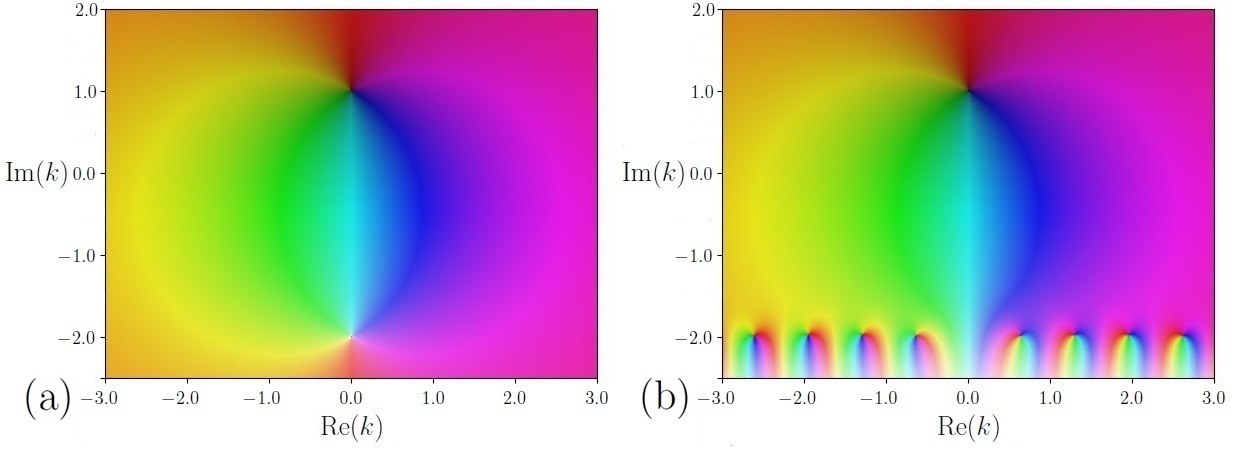}}
\caption{Complex plot of (a) the exact Jost function,~(\ref{exact_Jost_V1}) and (b) the Jost function of eq.~(\ref{f_calc_B0}) using $R$-matrix calculations on a Lagrange-Jacobi mesh, with $a=5$ and $N=40$, for potential 1 ($\ell=0$) with $\kappa_0=2$ and $\kappa_1=1$. The darkness or lightness of the colours are an indication of the modulus of the complex value and the hue of the colours is an indication of the argument. White dots represent poles and black dots represent zeros.}
\label{V1_Jost}
\end{figure}

\begin{figure}[htbp]
\centerline{\includegraphics[width=\textwidth]{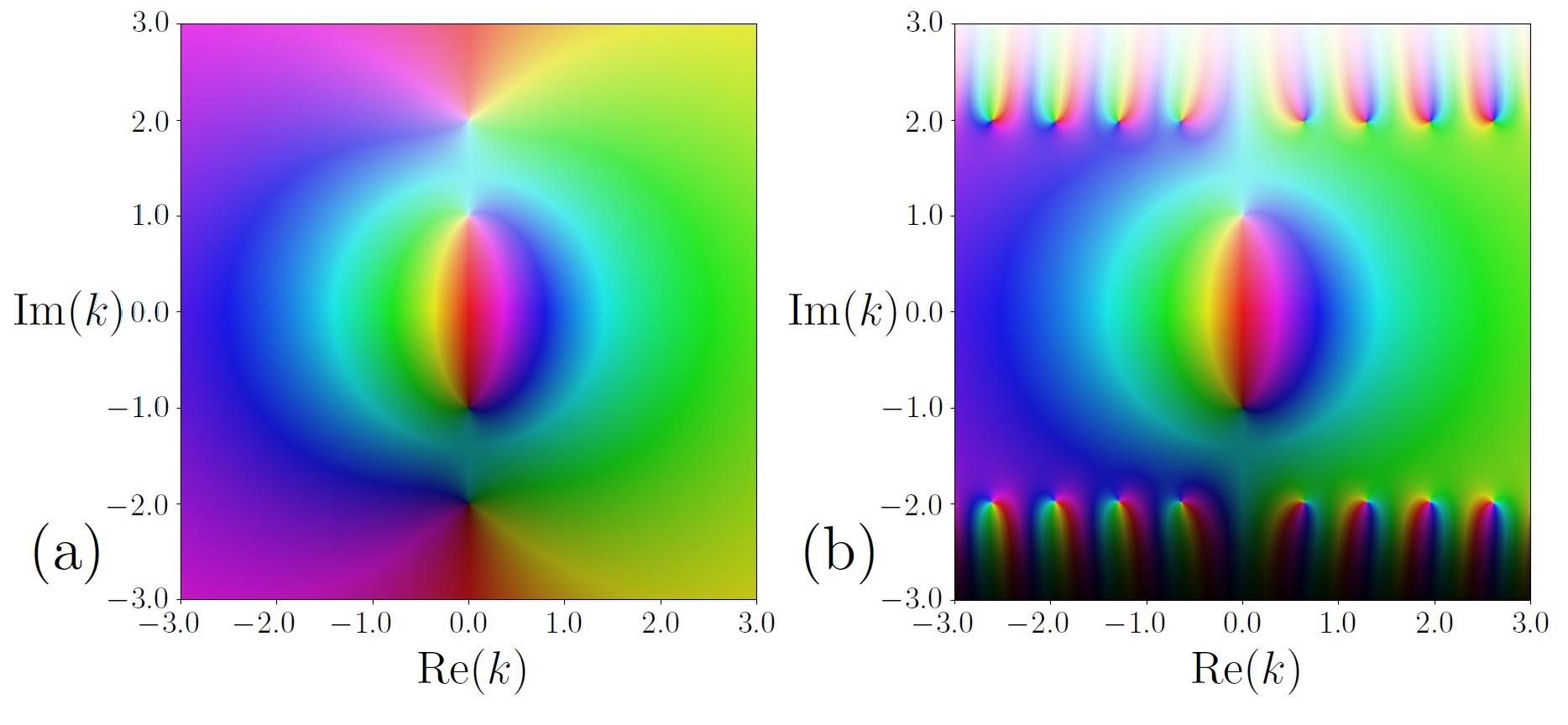}}
\caption{Complex plot of (a) the exact $S$-matrix and (b) the $S$-matrix from $R$-matrix calculations on a Lagrange-Jacobi mesh, with $a=5$ and $N=40$, for potential 1 ($\ell=0$) with $\kappa_0=2$ and $\kappa_1=1$. The darkness or lightness of the colours are an indication of the modulus of the complex value and the hue of the colours is an indication of the argument. White dots represent poles and black dots represent zeros.}
\label{V1_S}
\end{figure}

Figure \ref{V1_Jost} shows the exact Jost function, eq.~(\ref{exact_Jost_V1}) as well as the Jost function calculated from the $R$-matrix using eq.~(\ref{f_calc_B0}), in the complex $k$-plane. The most striking feature is the line of Siegert pseudostates that clearly appear around $k=-2i$ in the plot of the calculated Jost function. The pseudostates, which have regular spacing between them, replace the exact Jost-function pole at $k=-2i$, which does not appear in the calculated Jost function. Figure \ref{V1_S} shows the corresponding exact and calculated $S$-matrix using eq.~(\ref{S_k}), where the false $S$-matrix pole at $k=2i$ appears in the exact plot and the more complicated analytic structure is clearly seen.   

For increasing $a$, the number of Siegert pseudostates in an interval of $\mathrm{Re}(k)$ increases, and the spacing between states decreases. The Siegert pseudostates are thus strongly dependent on $a$. For increasing $N$, the accuracy of the Siegert pseudostates increases, until some point of convergence in the values for a number of digits is reached. In the vicinity of the pseudostates, specifically below $k=-2i$, the accuracy of the calculated Jost function is no longer reliable, which is also clear from table \ref{V1_table}.

Table \ref{V1_table} shows specific values of the calculated Jost function. Only values with $\mathrm{Re}(k) > 0$ are given, since the calculated and exact Jost functions are symmetrical around the real $k$-axis. If $\vert k \vert$ is large, the wave function becomes strongly oscillating and inaccuracies are to be expected for any numerical calculation. Values of $k$ with $-5 \le \mathrm{Re}(k) \le 5$ and $-5 \le \mathrm{Im}(k) \le 5$ are thus chosen. Apart from the Jost-function pole and Siegert pseudostates, the calculated Jost-function values are independent of $a$, up to an increasing accuracy with increasing $a$, provided that $N$ is big enough to reach convergence.  

\begin{table}[]
\centering
\begin{tabular}{ccccc}
\hline \hline
$k$        															& 
\multicolumn{2}{c}{Exact Jost function, eq.~(\ref{exact_Jost_V2})} 	& \multicolumn{2}{c}{Calculated Jost function, eq.~(\ref{f_calc_B0})} 	\\
\hline
  & $\mathrm{Re}[f_{0}(k)]$    & $\mathrm{Im}[f_{0}(k)]$    
  & $\mathrm{Re}[f_{0}(k)]$    & $\mathrm{Im}[f_{0}(k)]$             			\\
\hline  
$1.0+5.0i$ 															& 
$\textcolor{white}{+}5.8000000\times 10^{-1}$ 						&
$-6.0000000\times 10^{-2}$ 											& 
$\textcolor{white}{+}5.7983688\times 10^{-1}$ 						& $-6.0567545\times 10^{-2}$       									\\
$1.0+1.0i$ 															& 
$\textcolor{white}{+}1.0000000\times 10^{-1}$ 						& 
$-3.0000000\times 10^{-1}$ 											& 
$\textcolor{white}{+}1.0000001\times 10^{-1}$ 						& 
$-3.0000003\times 10^{-1}$   		  								\\
$1.0-1.5i$ 															& 
$-2.0000000\times 10^{-1}$ 											&
$-2.4000000\times 10^{\textcolor{white}{-1}}$							&
$-1.9149169\times 10^{-1}$ 											&
$-2.4231259\times 10^{\textcolor{white}{-1}}$\\
$1.0-2.5i$ 															& 
$\textcolor{white}{+}2.2000000\times 10^{\textcolor{white}{-1}}$ 		& 
$-2.4000000\times 10^{\textcolor{white}{-1}}$      					& 
$\textcolor{white}{+}3.9840112\times 10^{2\textcolor{white}{+}}$        					& $-2.4149833\times 10^{2\textcolor{white}{+}}$        									\\
$4.0+5.0i$ 															& 
$\textcolor{white}{+}6.7692308\times 10^{-1}$ 						& 
$-1.8461538\times 10^{-1}$ 											& 
$\textcolor{white}{+}6.7655990\times 10^{-1}$ 						& 
$-1.8409529\times 10^{-1}$    		   								\\
$4.0+1.0i$ 															& 
$\textcolor{white}{+}6.4000000\times 10^{-1}$ 						&
$-4.8000000\times 10^{-1}$ 											& 
$\textcolor{white}{+}6.4000002\times 10^{-1}$ 						& 
$-4.8000006\times 10^{-1}$   		    								\\
$4.0-1.5i$ 															& 
$\textcolor{white}{+}9.0769231\times 10^{-1}$ 						& 
$-7.3846154\times 10^{-1}$ 											& 
$\textcolor{white}{+}9.0057149\times 10^{-1}$ 						& 
$-7.3838099\times 10^{-1}$       									\\
$4.0-2.5i$ 															& 
$\textcolor{white}{+}1.0923077\times 10^{\textcolor{white}{-1}}$     	& $-7.3846154\times 10^{-1}$ 											&
$-1.2473599\times 10^{2\textcolor{white}{+}}$ 											& 
$-4.3576965\times 10^{\textcolor{white}{+1}}$  
																	\\
																	 \hline \hline
\end{tabular}%
\caption {Exact Jost-function values for various complex $k$ compared with values determined from $R$-matrix calculations on a Lagrange-Jacobi mesh with $a=5$ and $N=40$ for potential 1 ($\ell=0$) with $\kappa_0=2$ and $\kappa_1=1$.}
\label{V1_table}
\end{table}
The calculated Jost-function value at the exact Jost-function pole of $k=-2i$ is dependent on the choice of $a$. For a larger choice of $a$, the calculated value is larger, as shown in table~\ref{V1_pole_values}. The increase is approximately linear with increasing $a$. The calculated Jost function at $k=1+i$ and $k=1-2i$ are also given in table~\ref{V1_pole_values}, for reference. The first value is independent of $a$, up to an increasing accuracy with increasing $a$. The second value, which is in the vicinity of the pseudostates, fluctuates for increasing $a$, since the pseudostates also shift for increasing $a$. The value of $N$ must be increased with increasing $a$ in order for the calculated values to converge. For $a\ge 6$, the functions $I_{\ell}(ka,\eta)$ and $O_{\ell}(ka,\eta)$ oscillate strongly with respect to $k$. For $k=-2i$ in particular, the functions grow and shrink exponentially with respect to increasing $a$. This causes numerical difficulties, which can partly be addressed by using the asymptotic approximations for $I_{\ell}(ka,\eta)$ and $O_{\ell}(ka,\eta)$ given in Ref.~\cite{NIST}, for example. There are still exponentially increasing factors in eq.~(\ref{f_calc}), however, which is why convergence in the Jost-function value at $k=1-2i$ and $k=-2i$ is no longer reached even for a very large number of meshpoints, $N$.    

\begin{table}[]
\centering
\begin{tabular}{lllll}
\hline \hline
$a$   & $N$  & $f_{0}(1+i)$                     & $f_{0}(1-2i)$         & $f_{0}(-2i)$          \\
\hline
$2.8$ & $15$ & 
$0.10003\textcolor{white}{00000} -0.30006i$    &
$3.0987-0.0253i$ & 
$16.55$           \\
$3.2$ & $20$ & 
$0.100006\textcolor{white}{0000} -0.300000i$    &
$0.2918-0.6126i$ & 
$18.949$           \\
$3.6$ & $25$ & 
$0.100001\textcolor{white}{0000} -0.300001i$    &
$2.085\textcolor{white}{0}-9.911i$ & 
$21.350$           \\
$4.0$ & $30$ & 
$0.1000005\textcolor{white}{000} -0.3000005i$    &
$2.590\textcolor{white}{0}-3.813i$ & 
$23.750$           \\
$4.4$ & $37$ & 
$0.10000006\textcolor{white}{00} -0.30000004i$    &
$0.918\textcolor{white}{0}-6.142i$ & 
$26.150$           \\
$4.8$ & $39$ & 
$0.10000002\textcolor{white}{00} -0.30000003i$    &
$1.92\textcolor{white}{00}-6.57i$ & 
$28.55$           \\
$5.2$ & $44$ & 
$0.100000005\textcolor{white}{0}-0.300000005i$   & 
$4.20\textcolor{white}{00}-4.83i$ & 
$30.95$           \\
$5.6$ & $46$ & 
$0.100000002\textcolor{white}{0}-0.300000002i$   & 
$4.5\textcolor{white}{000}-2.0i$   & $33$              \\
$6.0$ & $53$ & $0.1000000000-0.3000000000i$ & 
does not converge & 
does not converge \\
\hline \hline
\end{tabular}%
\caption {Dependence on $a$ of Jost-function values determined from $R$-matrix calculations on a Lagrange-Jacobi mesh for potential 1 ($\ell=0$) with $\kappa_0=2$ and $\kappa_1=1$. The values at the exact Jost-function pole, $k=-2i$, increase as $a$ increases. The mesh size must be increased as $a$ increases, until the calculated values converge. Larger $N$ values affect the last digit of each given number.}
\label{V1_pole_values}
\end{table}
\begin{table}[]
\centering
\begin{tabular}{llllll}
\hline \hline
\multicolumn{2}{c}{Exact ($a=5$) Ref.~\cite{Baye2002}}             & \multicolumn{2}{c}{Eigenvalues of eq.~(\ref{siegert_method_lis0})}       &
\multicolumn{2}{c}{$f_{0}(k_{n0})=0$, eq.~(\ref{f_calc_B0})}          			\\ \hline
\multicolumn{1}{c}{$\mathrm{Re}(k_{n0})$} & \multicolumn{1}{c}{$\mathrm{Im}(k_{n0})$} & \multicolumn{1}{c}{$\mathrm{Re}(k_{n0})$} & \multicolumn{1}{c}{$\mathrm{Im}(k_{n0})$} & \multicolumn{1}{c}{$\mathrm{Re}(k_{n0})$} & \multicolumn{1}{c}{$\mathrm{Im}(k_{n0})$}
              \\ \hline
$0$                            & $\textcolor{white}{-}0.9999999999955$              & $0$                          & $\textcolor{white}{-}0.999999999995$               & $0$                          & $\textcolor{white}{-}0.999999999994$               \\
$0.6390266702$                 & $-1.9909403640$                & $0.6390184$                    & $-1.9909406$                  & $0.6390184$                    & $-1.99094068$                  \\
$1.2947020189$                 & $-1.9808654900$                & $1.2946844$                    & $-1.9808658$                   & $1.2946849$                    & $-1.9808660$                   \\
$1.9559919327$                 & $-1.9824365206$                & $1.9559654$                    & $-1.9824365$                   & $1.9559661$                    & $-1.9824368$                   \\
$2.6142071140$                 & $-1.9926145823$                & $2.6141719$                    & $-1.9926141$                   & $2.6141728$                    & $-1.9926145$                   \\
$3.2675033086$                 & $-2.0070370919$                & $3.2674595$                    & $-2.0070362$                   & $3.2674607$                    & $-2.0070367$                   \\
$3.9163165163$                 & $-2.0231436137$                & $3.9162643$                    & $-2.0231425$                   & $3.9162656$                    & $-2.0231430$                   \\
$4.5614908509$                 & $-2.0396289360$                & $4.5614302$                    & $-2.0396276$                   & $4.5614318$                    & $-2.0396283$                   \\
$5.2037975084$                 & $-2.0558538335$                & $5.2037285$                    & $-2.0558523$                   & $5.2037303$                    & $-2.0558533$                   \\
$5.8438509338$                 & $-2.0715187039$                & $5.8437734$                    & $-2.0715172$                   & $5.8437754$                    & $-2.0715184$                   \\
$6.4821222431$                 & $-2.0864975209$                & $6.4820362$                    & $-2.0864960$                   & $6.4820382$                    & $-2.0864976$                   \\
$7.1189698354$                 & $-2.1007533271$                & $7.1188752$                    & $-2.1007519$                   & $7.1188770$                    & $-2.1007536$                   \\
$7.7546673954$                 & $-2.1142944575$                & $7.7545640$                    & $-2.1142931$                   & $7.7545657$                    & $-2.1142949$                 \\
\hline \hline 
\end{tabular}%
\caption {Siegert state (the first line) and Siegert pseudostates, $k_{n\ell}$, of potential 1 ($\ell=0$) with $\kappa_0=2$ and $\kappa_1=1$, with a truncation at $a=5$. Results are from calculations on a Lagrange-Jacobi mesh with $N=40$. For comparison, exact values for a truncated potential are also given, obtained from solving for $k_{n\ell}$ in eq.~(49) of Ref.~\cite{Baye2002}, with $a=5$.}
\label{V1_siegert}
\end{table}

The Jost-function zero, which is a Siegert state corresponding to a bound state in this case, appears at $k=i$ for the calculated Jost function. Other than the pseudostates, it is independent of $a$, up to an increasing accuracy with increasing $a$, provided that $a$ is again chosen large enough for the potential to be approximately zero for $r \ge a$. Finding the $S$-matrix poles or calculating the Jost-function zeros to locate the bound state gives exactly the same result, as expected, with an accuracy of $8$ significant figures. 

The zeros of the Jost function were determined using a Newton-Raphson iterative method with various arbitrary starting values of $k$. Poles of the $S$-matrix were similarly determined using a Newton-Raphson method to find zeros of $1/S_{\ell}(k)$. However, the calculated $S$-matrix has a more complicated structure than the calculated Jost function, as seen in figure~\ref{V1_S}. When using the Newton-Raphson method to find the zeros of the inverse $S$-matrix, starting values need to be chosen much closer to the actual zeros in order for the method to converge. Finding the Jost-function zeros proved to be easier, as the Newton-Raphson method converged quickly for most starting values.                      

Apart from the bound state, which is the Siegert state, the Siegert pseudostates were also calculated by finding the zeros of the calculated Jost function. Table \ref{V1_siegert} gives a few of the Siegert pseudostates obtained by solving the eigenvalue eq.~(\ref{siegert_method_lis0}) using the algebraic method of Ref.~\cite{Tolstikhin1998}, with the results from finding the zeros of the calculated Jost function, eq.~(\ref{f_calc_B0}). Both methods give similar accuracy when compared to the Lagrange-Legendre mesh calculations of Ref.~\cite{Baye2002}, where $a=5$ and $N=25$. However, accuracy is better for the choice of $a=5$ and $N=40$ in the current work, due to the larger number of mesh points. The values of Ref.~\cite{Baye2002} given in table \ref{V1_siegert} were calculated by solving eq.~(49) of that paper. This equation is obtained from the approximate behaviour of the wave functions of the Siegert pseudostates for the Bargmann potential truncated at $r=a$, and is only applicable to situations where $\ell=0$. The simple iterative $R$-matrix procedure where $B=L_{\ell}(k)$, described in Section \ref{section-R-matrix_siegert}, was also attempted to calculate the Siegert pseudostates, without success. 

A virtual state can be created with potential 1 if $\kappa_1<0$. In particular, the values $(\kappa_0,\kappa_1)=(2,-0.5)$ and $(\kappa_0,\kappa_1)=(2,-1)$ have been tested. The Siegert method, determining eigenvalues, $k_{n\ell}$, of eq.~(\ref{siegert_method_lis0}), gives almost identical results to finding the zeros of the Jost function using eq.~(\ref{f_calc_B0}). These values are within an accuracy of around $5$ digits. The simple $R$-matrix iterative method setting $B=L_{\ell}(k)$ also gives good results, but convergences only occurs after a large number of iterations. 

In conclusion, finding the zeros of the Jost function consistently gives readily converging results with good accuracy for both Siegert states (bound and virtual) and Siegert pseudostates.

\subsection{Potential 2}
\begin{figure}[htbp]
\centerline{\includegraphics[width=\textwidth]{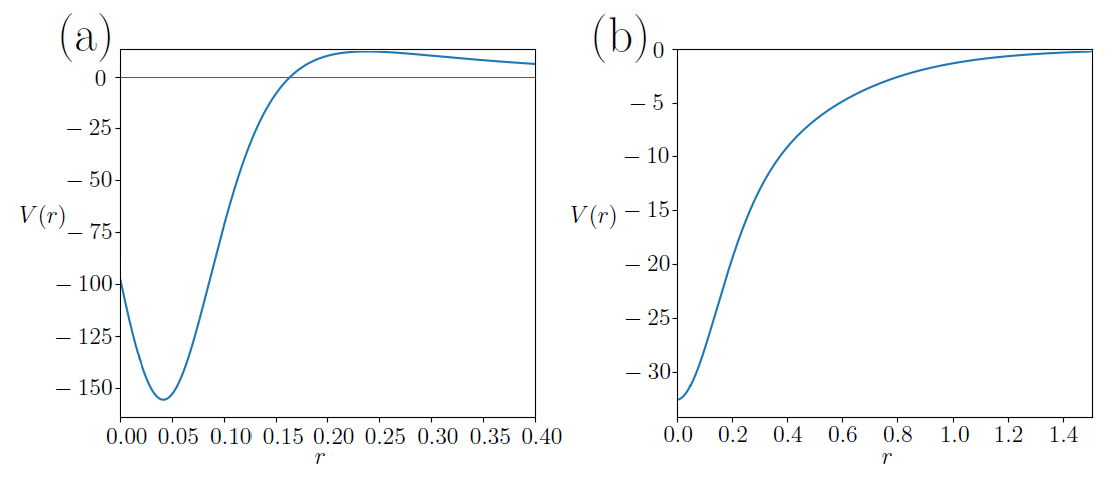}}
\caption{Plots of (a) potential 2, eq.~(\ref{V2}), and (b) potential 3, eq.~(\ref{V3}), with $\kappa_0=1.5$, $\kappa_1=9.75$ and $\alpha=0.1+0.5i$. Note the difference in scales.}
\label{V2_V3_plots}
\end{figure}

Parameters for potential 1 were chosen to correspond with the Bargmann potential used in Ref.~\cite{Baye2002}, so that results could easily be compared. No other study (to our knowledge) has used potential 2 and 3 in calculations. The parameters, $\alpha$, $\kappa_0$ and $\kappa_1$, are chosen to be the same for both potential 2 with $\ell=0$ and potential 3 with $\ell=1$. This means they will have exactly the same Jost function, $S$-matrix, resonance, and Jost-function poles. This simplifies the comparison of results.

The parameter, $\alpha$, is firstly chosen to result in a relatively narrow resonance occurring at an experimentally attainable energy. The main restriction is then condition (\ref{V3_cond}) of potential 3. For the resonance to have a positive real part of the energy,  $\alpha_R < \alpha_I$. Also, $\kappa_0$, $\kappa_1$, $\alpha_R$ and $\alpha_I$ must be positive. Furthermore, it is known that the calculated Jost function is inaccurate in the vicinity of an exact Jost-function pole, from the results of potential 1. If $\alpha_R$ is chosen too close to $\kappa_0$ and $\kappa_1$, we will not be able to accurately locate the resonance at $k=-i\alpha$ with the $R$-matrix calculations.      

A choice of $\kappa_0 = 1.5$, $\kappa_1= 0.75$ and $\alpha = 0.2+0.4i$ satisfies all the requirements, for example, and we were able to calculate the Jost function from the $R$-matrix for such a choice. However, better accuracy for the resonance calculation was obtained with larger values of $\kappa_1$, so that the Jost-function poles are further from the resonance. Due to the condition (\ref{V3_cond}), a larger $\kappa_1$ requires an adjustment in the choice of $\alpha$ as well.  

The results for potential 2, eq.~(\ref{V2}), for $\kappa_0=1.5$, $\kappa_1=9.75$ and $\alpha=0.1+0.5i$ are reported here. The potential has a value of $V(0)=-97.7925$ at the origin for this choice. Figure \ref{V2_V3_plots} shows plots of both potential 2 and potential 3 for the same parameters.

The Jost-function values at various complex $k$ are given in table \ref{V2_table}, where the calculated Jost-function values were obtained with $a=5$ and $N=70$.

\begin{figure}[htbp]
\centerline{\includegraphics[width=\textwidth]{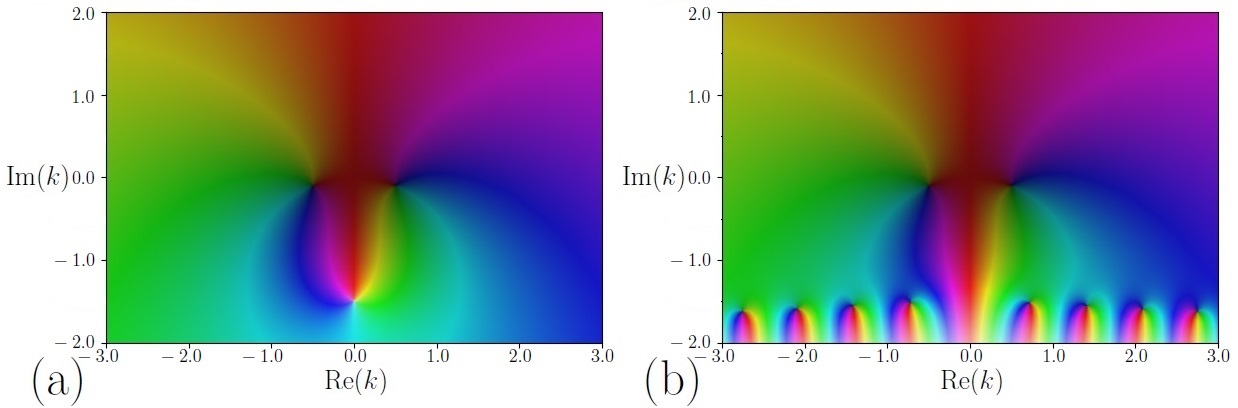}}
\caption{Complex plot of (a) the exact Jost function,~(\ref{exact_Jost_V1}) and (b) the Jost function of eq.~(\ref{f_calc_B0}) using $R$-matrix calculations on a Lagrange-Jacobi mesh with $a = 5$ and $N = 50$, for potential 2 (and very similar for potential 3). The darkness or lightness of the colours are an indication of the modulus of the complex value and the hue of the colours is an indication of the argument. White dots represent poles and black dots represent zeros.}
\label{V2_Jost}
\end{figure}

Figure \ref{V2_Jost} shows the exact Jost function with the calculated Jost function for potential 2, where Siegert pseudostates around the first pole with $k=-1.5i$ are again present for the calculated Jost function. The calculated Jost-function values near the Siegert pseudostates are, once again, not accurate and the Jost-function pole is once again absent in the calculated Jost function.

The calculated Jost function also becomes very unstable for values of $\mathrm{Im}(k)$ lower than the first pole at $k=-i\kappa_0$, the further one moves from the pole along the negative $\mathrm{Im}(k)$ axis. The plots of Figure \ref{V2_Jost} are thus scaled so that the second pole with $k=-9.75i$ is not visible, due to the instability in the calculated Jost function in this region. 

There is agreement with the Siegert pseudostates calculated from finding the eigenvalues of eq.~(\ref{siegert_method_lis0}) and by finding zeros of eq.~(\ref{f_calc_B0}). The first few Siegert pseudostates are given in table \ref{V2_siegert}.

\begin{table}[]
\centering
\begin{tabular}{ccccc}
\hline \hline
$k$        															& 
\multicolumn{2}{c}{Exact Jost function, eq.~(\ref{exact_Jost_V2})} 	& \multicolumn{2}{c}{Calculated Jost function, eq.~(\ref{f_calc_B0})} 	\\
\hline
  & $\mathrm{Re}[f_{0}(k)]$    & $\mathrm{Im}[f_{0}(k)]$    
  & $\mathrm{Re}[f_{0}(k)]$    & $\mathrm{Im}[f_{0}(k)]$             			\\
\hline  
$1.0+5.0i$ 															& 
$\textcolor{white}{+}2.7645615\times 10^{-1}$ 						&
$-4.5589531\times 10^{-2}$ 											& 
$\textcolor{white}{+}2.7664388\times 10^{-1}$ 						& $-4.4383169\times 10^{-2}$       									\\
$1.0+1.0i$ 															& 
$\textcolor{white}{+}4.8578349\times 10^{-2}$ 						& 
$-6.0148285 \times 10^{-2}$ 											& 
$\textcolor{white}{+}4.8576235\times 10^{-2}$ 						& 
$-6.0145961 \times 10^{-2}$   		  								\\
$1.0-1.5i$ 															& 
$-3.5200000 \times 10^{-1}$ 											&
$\textcolor{white}{+}1.0400000 \times 10^{-1}$		&
$-5.1496562 \times 10^{-1}$ 											&
$\textcolor{white}{+}3.2618065 \times 10^{-1}$		\\
$1.0-2.5i$ 															& 
$-6.6588098 \times 10^{-1}$ 						& 
$-7.7362894 \times 10^{-2}$      					& 
$-4.2880295 \times 10^{3\textcolor{white}{+}}$        									& 
$\textcolor{white}{+}8.8607251 \times 10^{2\textcolor{white}{+}}$        					\\
$4.0+5.0i$ 															& 
$\textcolor{white}{+}3.1514242 \times 10^{-1}$ 						& 
$-1.7543530 \times 10^{-1}$ 											& 
$\textcolor{white}{+}3.1614624 \times 10^{-1}$ 						& 
$-1.7618833 \times 10^{-1}$    		   								\\
$4.0+1.0i$ 															& 
$\textcolor{white}{+}1.0531244 \times 10^{-1}$ 						&
$-2.9594854 \times 10^{-1}$ 											& 
$\textcolor{white}{+}1.0530816 \times 10^{-1}$ 						& 
$-2.9593692 \times 10^{-1}$   		    								\\
$4.0-1.5i$ 															& 
$-1.1075093 \times 10^{-1}$ 											& 
$-4.7157621 \times 10^{-1}$ 											& 
$-4.0005488 \times 10^{-2}$ 											& 
$-4.2100557 \times 10^{-1}$       									\\
$4.0-2.5i$ 															& 
$-2.1254330 \times 10^{-1}$  											& $-5.9726527 \times 10^{-1}$ 											&
$\textcolor{white}{+}1.1328721 \times 10^{3\textcolor{white}{+}}$ 						& 
$\textcolor{white}{+}1.2653667 \times 10^{3\textcolor{white}{+}}$  
																	\\
																	 \hline \hline
\end{tabular}%
\caption {Exact Jost-function values for various complex $k$ compared with values determined from $R$-matrix calculations on a Lagrange-Jacobi mesh with $a=5$ and $N=70$ for potential 2 ($\ell=0$) with $\kappa_0=1.5$, $\kappa_1=9.75$ and $\alpha = 0.1+0.5i$.}
\label{V2_table}
\end{table}

\begin{table}[]
\centering
\begin{tabular}{llll}
\hline \hline
\multicolumn{2}{c}{Eigenvalues of eq.~(\ref{siegert_method_lis0})}       	&
\multicolumn{2}{c}{$f_{0}(k_{n0})=0$, eq.~(\ref{f_calc_B0})}          			\\ \hline
\multicolumn{1}{c}{$\mathrm{Re}(k_{n0})$} & \multicolumn{1}{c}{$\mathrm{Im}(k_n)$} & \multicolumn{1}{c}{$\mathrm{Re}(k_{n0})$} & \multicolumn{1}{c}{$\mathrm{Im}(k_{n0})$}
              \\ \hline
$0.4999960993$ & $-0.0999989021$ & $0.4999960990$ & $-0.0999989094$ \\
$0.7193780438$ & $-1.5179751744$ & $0.7193779983$ & $-1.5179750909$ \\
$1.4153380880$ & $-1.5551524081$ & $1.4153380062$ & $-1.5551522487$  \\
$2.0916341936$ & $-1.5927307397$ & $2.0916340745$ & $-1.5927305053$  \\
$2.7565524486$ & $-1.6260844633$ & $2.7565522952$ & $-1.6260841477$  \\
$3.4145388619$ & $-1.6552343620$ & $3.4145386852$ & $-1.6552339645$  \\
$4.0678385625$ & $-1.6808440710$ & $4.0678383661$ & $-1.6808435958$  \\
$4.7177133517$ & $-1.7035408074$ & $4.7177131405$ & $-1.7035402554$  \\
$5.3649460996$ & $-1.7238391612$ & $5.3649458758$ & $-1.7238385401$  \\
$6.0100602871$ & $-1.7421515475$ & $6.0100600458$ & $-1.7421508630$  \\
\hline \hline 
\end{tabular}%
\caption {Siegert state (the first value) and Siegert pseudostates, $k_{n\ell}$, of potential 2 ($\ell=0$)  with $\kappa_0=1.5$, $\kappa_1=9.75$ and $\alpha = 0.1+0.5i$. Results are from $R$-matrix calculations on a Lagrange-Jacobi mesh with $a=5$ and $N=70$.}
\label{V2_siegert}
\end{table}

The characteristic line of pseudostates that appear around the exact Jost-function pole for potential 1, appears around the first pole of potential 2. However, neither the Jost method nor the Siegert method of finding the eigenvalues of eq.~(\ref{siegert_method_lis0}) gives any distinct line of pseudostates around the second exact Jost-function pole for potential 2. No pseudostates appear around the second pole for parameters where the poles are closer together and nearer the origin, either, such as $\alpha = 0.2+0.4i$, $\kappa_0 = 1.5$ and $\kappa_1= 0.75$. This is because the calculated Jost function is not accurate below the first pole, in the $k$-plane.   

The exact resonance is located at $k=-i\alpha$. As can be seen in the first value of table \ref{V2_siegert}, very similar results with an accuracy of $5$ digits was obtained for $a=5$ and $N=70$ with the Siegert method, where the eigenvalues of (\ref{siegert_method_lis0}) are calculated, and by finding the zeros of the calculated Jost function, eq.~(\ref{f_calc_B0}). Similar accuracy was obtained with the iterative $R$-matrix method, letting $B=L_{\ell}(k)$. 

There is once again a strong dependence of the calculated Jost function on the channel radius, $a$, at an exact Jost-function pole. The calculated Jost-function value becomes larger as $a$ is increased, but, as with potential 1, the value does not converge for larger $a$ due to the inaccuracy around the nearby pseudostates.  

\subsection{Potential 3}
The $\ell=1$ potential with resonance, eq.~(\ref{V3}), is used with the same parameters as potential 2: $\kappa_0=1.5$, $\kappa_1=9.75$ and $\alpha=0.1+0.5i$, giving $V(0)=-32.5975$. The potential is plotted in Figure \ref{V2_V3_plots}, with potential 2. The results for the Jost functions at various complex $k$ are given in table \ref{V3_table}. For these results, $a=5$ and $N=50$ in the $R$-matrix calculations.

\begin{table}[]
\centering
\begin{tabular}{ccccc}
\hline \hline
$k$        															& 
\multicolumn{2}{c}{Exact Jost function, eq.~(\ref{exact_Jost_V2})} 	& \multicolumn{2}{c}{Calculated Jost function, eq.~(\ref{f_calc_B0})} 	\\
\hline
  & $\mathrm{Re}[f_{1}(k)]$    & $\mathrm{Im}[f_{1}(k)]$    
  & $\mathrm{Re}[f_{1}(k)]$    & $\mathrm{Im}[f_{1}(k)]$             			\\
\hline  
$1.0+5.0i$ 															& 
$\textcolor{white}{+}2.7645615\times 10^{-1}$ 						&
$-4.5589531\times 10^{-2}$ 											& 
$\textcolor{white}{+}2.7327711\times 10^{-1}$ 						& $-5.6378963\times 10^{-2}$       									\\
$1.0+1.0i$ 															& 
$\textcolor{white}{+}4.8578349\times 10^{-2}$ 						& 
$-6.0148285 \times 10^{-2}$ 											& 
$\textcolor{white}{+}4.8578173 \times 10^{-2}$ 						& 
$-6.0147997  \times 10^{-2}$   		  								\\
$1.0-1.5i$ 															& 
$-3.5200000 \times 10^{-1}$ 											&
$\textcolor{white}{+}1.0400000 \times 10^{-1}$		&
$-5.5927914  \times 10^{-1}$ 											&
$\textcolor{white}{+}3.2277235  \times 10^{-1}$		\\
$1.0-2.5i$ 															& 
$-6.6588098 \times 10^{-1}$ 						& 
$-7.7362894 \times 10^{-2}$      					& 
$-4.9487867  \times 10^{3\textcolor{white}{+}}$        									& 
$\textcolor{white}{+}7.4163770  \times 10^{2\textcolor{white}{+}}$        					\\
$4.0+5.0i$ 															& 
$\textcolor{white}{+}3.1514242 \times 10^{-1}$ 						& 
$-1.7543530 \times 10^{-1}$ 											& 
$\textcolor{white}{+}3.0789459  \times 10^{-1}$ 						& 
$-1.6628377  \times 10^{-1}$    		   								\\
$4.0+1.0i$ 															& 
$\textcolor{white}{+}1.0531244 \times 10^{-1}$ 						&
$-2.9594854 \times 10^{-1}$ 											& 
$\textcolor{white}{+}1.0531189  \times 10^{-1}$ 						& 
$-2.9594694  \times 10^{-1}$   		    								\\
$4.0-1.5i$ 															& 
$-1.1075093 \times 10^{-1}$ 											& 
$-4.7157621 \times 10^{-1}$ 											& 
$-2.6737520  \times 10^{-2}$ 											& 
$-4.0030798  \times 10^{-1}$       									\\
$4.0-2.5i$ 															& 
$-2.1254330 \times 10^{-1}$  											& $-5.9726527 \times 10^{-1}$ 											&
$\textcolor{white}{+}1.3064955  \times 10^{3\textcolor{white}{+}}$ 						& 
$\textcolor{white}{+}1.6776195  \times 10^{3\textcolor{white}{+}}$  
																	\\
																	 \hline \hline
\end{tabular}%
\caption {Exact Jost-function values for various complex $k$ compared with values determined from $R$-matrix calculations on a Lagrange-Jacobi mesh with $a=5$ and $N=50$ for potential 3 ($\ell=1$) with $\kappa_0=1.5$, $\kappa_1=9.75$ and $\alpha = 0.1+0.5i$.}
\label{V3_table}
\end{table}

The plot of the exact Jost function compared with the calculated Jost function for potential 3 is almost identical to that of potential 2, given in Figure \ref{V2_Jost}. This is to be expected, since $a=5$ for the calculations with both potentials. For this reason, the plot of the Jost functions for potential 3 is not shown.     

As with potential 1 and 2, the calculated Jost-function values near the Siegert pseudostates are not accurate and the exact Jost-function pole is absent in the calculated Jost function. As before, there is a strong dependence of the calculated Jost function on the channel radius, $a$, at the exact Jost-function pole. For increasing $a$, the calculated Jost-function value at the exact Jost-function pole becomes larger and pseudostates are closer together. 

The Siegert pseudostates cannot be determined from finding the eigenvalues of eq.~(\ref{siegert_method_lis0}), since $\ell \ne 0$ in this case. They could be determined by finding zeros of the calculated Jost function, and the first few are given in table \ref{V3_siegert}.     

The exact resonance is also given by $k=-i\alpha$. An accuracy of $5$ digits was obtained for $a=5$ and $N=50$ by finding the zeros of the calculated Jost function (the value also appears in table \ref{V3_siegert}), with similar accuracy from the iterative $R$-matrix method. In fact, this accuracy is obtained with both methods for $N=50$, which is an improvement on the situation for potential 2, where a larger number of mesh points is necessary. This is because potential 2 has a large negative value at the origin, requiring a larger mesh to reflect the sharply-changing behaviour of the potential accurately. 

There is a small but significant difference in the Siegert pseudostates for potential 2 and 3, as can be seen when comparing the values of tables \ref{V2_siegert} and \ref{V3_siegert}. It is known that the Siegert pseudostates depend on $a$, which is chosen to be the same for both potentials. The choice of $N$ does not significantly affect the Siegert pseudostates, provided it is large enough. The difference is thus due to the $\ell$ dependence.     

\begin{table}[]
\centering
\begin{tabular}{ll}
\hline \hline
\multicolumn{2}{c}{$f_{1}(k_{n1})=0$, eq.~(\ref{f_calc_B0})}          
              \\ \hline
\multicolumn{1}{c}{$\mathrm{Re}(k_{n1})$} & \multicolumn{1}{c}{$\mathrm{Im}(k_{n1})$}
              \\ \hline
$0.4999993123$                            & $-0.0999990224$                 \\
$0.7077530379$                 & $-1.5112345900$                   	\\
$1.4020996661$                 & $-1.5398733315$                   	\\
$2.0799596802$                 & $-1.5727489640$                  	\\
$2.7466118809$                 & $-1.6036327919$                   	\\
$3.4060236455$				  & $-1.6313943967$						\\
$4.0604426993$				  & $-1.6561677428$						\\
$4.7111989944$				  & $-1.6783307619$						\\
$5.3591364555$				  & $-1.6982724970$						\\
$6.0048238958$				  & $-1.7163377041$						\\			
\hline \hline 
\end{tabular}%
\caption {Siegert state (the first value) and Siegert pseudostates, $k_{n\ell}$, of potential 3 ($\ell=1$) with $\kappa_0=1.5$, $\kappa_1=9.75$ and $\alpha=0.1+0.5i$, obtained by finding zeros of the Jost function from $R$-matrix calculations on a Lagrange-Jacobi mesh with $a=5$ and $N=50$.}
\label{V3_siegert}
\end{table}

\section{Conclusion}
A method for determining the Jost function from $R$-matrix calculations on a Lagrange-Jacobi mesh for any partial wave is developed. The focus of the study is to analyse the behaviour of this calculated Jost function in the complex $k$-plane, specifically its relation to the Siegert states and Siegert pseudostates. 

The accuracy of the calculated Jost function is, firstly, a reflection of the reliability of the $R$-matrix method since inaccuracies in the calculated Jost function are due to inaccuracies in the $R$-matrix.
The zeros of the calculated Jost function correspond with the Siegert states and the Siegert pseudostates. Poles on the negative imaginary axis of $k$ that may exist for the exact Jost function, do not appear in the calculated Jost function. Poles nearest to the origin are replaced by a distinct horizontal line of equally-spaced Siegert pseudostates that occur around the point where the pole should be. The calculated Jost function is reliable for all $k$ (provided $|k|$ is not too large) apart from intervals around the Siegert pseudostates, specifically below the position of an exact Jost-function pole.

A further important part of this study is to accurately determine the values of $k$ that correspond with Siegert states (bound, virtual or resonance states) and Siegert pseudostates. Three methods are used. The first method, proposed here, is by finding the zeros of the Jost function calculated with the $R$-matrix on a Lagrange-Jacobi mesh. This method is simple to apply and gives excellent results for all the test potentials, and there is no restriction on its application.

Finding the zeros of the calculated Jost function is identical to finding the poles of the $S$-matrix from $R$-matrix calculations. The $S$-matrix can accurately be determined with the $R$-matrix on various meshes, and in practical calculations the poles can be determined in various ways. However, it is found that the Newton-Raphson procedure used to find Jost-function zeros in this study is more robust in finding Siegert states and Siegert pseudostates than trying to locate $S$-matrix poles.
 
The second method is the Siegert method of Ref.~\cite{Tolstikhin1997,Tolstikhin1998}, which is only applicable to $s$-wave scattering but gives good results where it can be applied. The third method is the simple iterative $R$-matrix scheme proposed in Ref.~\cite{Descouvemont2010}. When it converges, which is not always guaranteed, it gives very accurate results. Siegert pseudostates could not be determined with this method. The reason for this lack of convergence of the Siegert pseudostates may be because they are mathematically similar to wide resonances, which are numerically difficult to locate.     

\section*{Acknowledgements} 
This work has received funding from the F.R.S.-FNRS under Grant No.~4.45.10.08. as well as from a bilateral grant of the South African National Research Foundation (NRF) and F.R.S.-FNRS, Grant PINT-BILAT-M R.M004.19 ``Multi-channel quantum resonances''.

We also extend our gratitude to Nicholas Trofimenkoff who contacted one of the authors to point out errors in Ref.~\cite{Baye2014}, which have been corrected here.

\end{document}